\documentclass[conference]{IEEEtran}
\IEEEoverridecommandlockouts
\usepackage{cite}
\usepackage{amsmath,amssymb,amsfonts}
\usepackage{algorithmic}
\usepackage{graphicx}
\usepackage{textcomp}
\usepackage{xcolor}
\def\BibTeX{{\rm B\kern-.05em{\sc i\kern-.025em b}\kern-.08em
    T\kern-.1667em\lower.7ex\hbox{E}\kern-.125emX}}
\usepackage{xspace}
\usepackage{url}
\newcommand\srs{SR system\xspace}
\newcommand\srss{SR systems\xspace}
\begin{document}

\title{Preparing for Super-Reactivity:  \\
Early  Fault-Detection 
in the Development of Exceedingly Complex Reactive Systems
}

\author{\IEEEauthorblockN{David Harel}
\IEEEauthorblockA{\textit{Dept. of Computer Science and Applied Mathematics} \\
\textit{Weizmann Institute of Science}\\
Rehovot, Israel \\
 david.harel@weizmann.ac.il}
 \and
 \IEEEauthorblockN{Assaf Marron}
 \IEEEauthorblockA{\textit{Dept. of Computer Science and Applied Mathematics} \\
 \textit{Weizmann Institute of Science}\\
 Rehovot, Israel \\
 assaf.marron@weizmann.ac.il}
 }

\maketitle

\begin{abstract}
We introduce the term \emph{Super-Reactive Systems} to refer to reactive systems whose construction and behavior are complex, constantly changing and evolving, and heavily interwoven with other systems and the physical world. 
Finding hidden faults in such systems early in planning and development is critical for human safety, the environment, society and the economy.
However, the  complexity of the system and its interactions and the absence of adequate technical details pose a great obstacle.
We propose an architecture for models and tools to overcome such barriers and  enable simulation, systematic analysis, and fault detection and handling, early in the development of super-reactive systems. 
The approach is facilitated  by  the inference and abstraction capabilities and the power and knowledge afforded by large language models and associated AI tools. It is based on: (i) deferred, just-in-time interpretation of model elements that are stored in natural language form, and (ii) early capture of tacit interdependencies among seemingly orthogonal requirements. 

\end{abstract}


\section{Introduction}\label{sec:introduction}

Since the 1985 identification of the category of \emph{reactive systems}\cite{harelPnueli1984ReactiveSystems}, a plethora of methods, languages and tools have been introduced to support the development of such systems. 
Today, complex reactive systems are penetrating almost every aspect of life, including communications, commerce, finance, healthcare, aviation, land transportation, manufacturing, and more. The complexity of new systems is compounded by the fact that they are interwoven with other systems and with the physical world, and are constantly changing and evolving. We term this emerging kind of system \emph{super-reactive} (SR). 
While system and software engineering (SySE) is benefitting from new developments in generative AI and large language models (LLMs), the challenge of building safe and reliable \srss remains open. Despite applying the best tools and methodologies, any given system is likely to conceal undesired and very often even unsafe behaviors and impending failures, with the risk of adverse effects on human life, the environment, society and the economy. Thus, while early discovery and handling of such faults is required, it remains a tantalizing challenge.  In this paper, we propose tackling this issue relying on the following principles: (1) Allow model elements that are expressed in natural language (NL), benefitting from the expressive power of natural language, its sensitivity to delicate context variations and its ability to navigate multiple levels of abstraction, and carrying out just-in-time (JIT), deferred, interpretation of such NL elements. 
(2) Discover and document otherwise-tacit interdependencies among separately specified, seemingly orthogonal requirements. 

A key enabler for our approach is the availability of large language models and other AI tools, the power and breadth of which is ever-growing. We exploit these in novel ways, in order to render possible principles (1) and (2) above.


\begin{figure*}[t] 
	\centering
\includegraphics[scale=0.55]{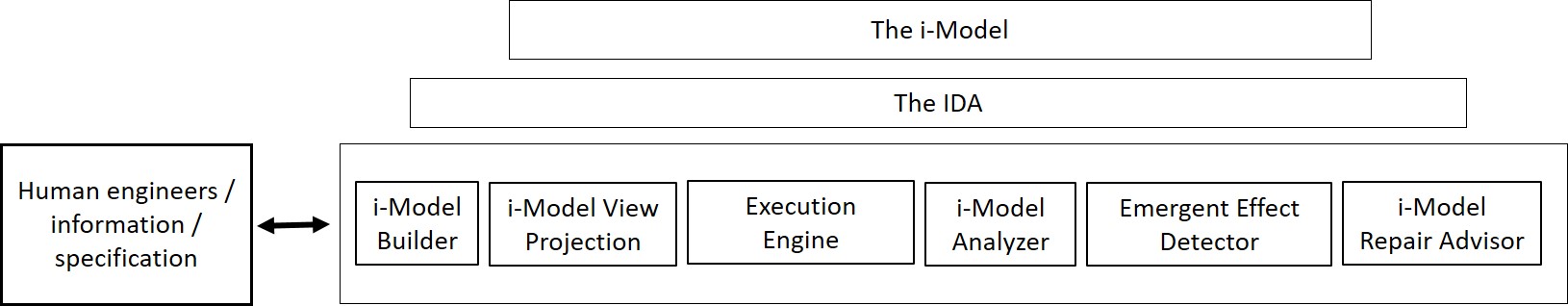}
	\caption{Solution architecture blueprint. See explanation in text.}
 \label{fig:imodelSolutionArchitecture}
\end{figure*}

\section{The Problem}

As the complexity and pervasiveness of reactive systems and systems of systems keeps growing, so do the risks associated with hidden faults: 
potential failure points, malfunction, undesired emergent behaviors and absence of desired ones. 
Recent well-known cases involving actual system code, maintenance procedures, interfaces with  other systems and with humans, etc., include: the Crowdstrike server failures in July 2024, the 2023 accident in which a robotic taxi hit a pedestrian in San Francisco, the failure of the USA FAA notification system in 2023, and the crashes of 737 MAX airplanes. 
Similar kinds of problems obviously occur frequently without gaining broad attention. 
Furthermore, beyond such direct effects, issues with an \srs may inflict excessive rigidity and regulation on the behavior of humans and of other systems, in an effort to accommodate the system's limitations. Compare for example the assignment of cars to driving sides and lanes on roads; one would like to avoid having such restrictions enforced in sidewalks and hallways shared by humans and robots as an emergency response to an unanticipated problematic reality.   

Published discussion of such problems call for early assessment and preemptive technical, economic, and regulatory activities. Early in the days of model-based system engineering (MBSE), France and Rumpe wrote: \emph{It is our view that software engineering is inherently a modeling
activity, and that the complexity of software will overwhelm
our ability to effectively maintain mental models of a system.}~\cite{franceRumpe2007MDEresearchroadmap}. 
Over the years there was great progress in the ability to build executable models. Examples include
UML, SysML, Rhapsody, STATEMATE, fUML, xUML, Ptolemy, MATLAB with Simulink and Stateflow, SCADE,  UPPAAL, BPMN, Arcadia, Cameo and others; see also list of SySE tools in  ~\cite[pp.76,79,208]{laplante2022SWEngineering}).
In parallel, there were significant advances in applying formal methods to such models~\cite{palanque2017SOTAinFM,fremontVincentelli2023scenic,saqui2021ModelCheckingSysMLCockpitDoor,zahid2022systematicMappingFMinRE,weyersPalanque2017handbook,huangMcGinnis2020verifyingSysMLPetriNets,Rahim2021HierachicalpETRINETSsysmlVerActivityReqRelations,Ghezzi2020earlyValidationMultiConcerns,harelMarronMizrahiWeissKatzKantor2013compositional}. 
And, many organizations developed elaborate ad-hoc models to help study the systems early in, and throughout, development~\cite{Lattimore2022EuropaSalnderSImulation,gorecki2019riskModelingIndustrial,lo2021digitalTwinInDev}. 

However, despite such advances, ensuring the safety and correctness of complex systems is still an unsolved problem. 
For example, in a 2024 workshop on safety of autonomous transportation (summarized in~\cite{dagstuhl2024AVsafety}) many open SySE issues and challenges were discussed, including: (i) incorporating general and domain knowledge in testing and verification; (ii) ensuring that ML training data covers rare but critical scenarios; (iii) exhaustively covering all possible interactions; and, (iv) enhancing usability of formal methods. Similar conclusions about gaps in present methods for early issue identification appear in \cite{cederbladh2024earlyVandVofModels,
horvath2023pragmaticVerificationWIthExecutableSysML,
harelSifakisMarron2020autonomicsPNAS,EdwardLee2024certaintyOrIntelligence}.
Given that uncovering hidden faults in well-specified and even fully developed systems is still an open problem, it is evident that preemptive fault discovery in super-reactive systems (e.g.,  complex systems of systems interwoven with their environment) at early development stages poses a major challenge.
The added difficulty stems in part from the informal and imprecise nature of requirements in early development stages, the limited scalability of the tools, and reliance on engineers to infer at development time undocumented relations among separately specified requirements that use different abstractions and a variety of terminologies, the dependence of verification on translation to state machines and Petri nets, and the absence of executable and analyzable semantics of certain specification artifacts.  

Current AI-based solutions  assist in various activities of development, including code generation to debugging, with a prominent example being the GitHub Copilot; see, also, e.g.,~\cite{AIESE2024Conf} and references therein.  However, applying such tools in the context of early specification is mostly limited to automated modeling, discussed in Section~\ref{sec:iModelBuilder}. 

The roadmap presented here addresses these issues by enabling rigorous execution and analysis subject to domain expertise and world knowledge, and doing so at higher levels of abstraction. The approach relies on the ability of AI-based tools to mimic humans' flexible navigation of complex abstraction relations using natural language, and will extend the present use of abstraction as in object-oriented inheritance relations, and in counter-example guided abstraction refinement (CEGAR) in formal verification~\cite{clarke2003counterexamples,seippHelmert2018CegarAndAbstractionTypesPlanning}.

\section{The roadmap}

Below we list the elements of an approach and an architecture for modeling \srss, including a set of 
intelligent tools for simulation and analysis, which,  together, can enable the much desired early preemptive discovery of hidden faults, while 
coping with the existing challenges and technology gaps. 

In bounding the scope of the proposed solution,  we exclude 
the use of AI, ML and LLMs in runtime decision making, monitoring, development operations (DevOps), or the formal verification of final 
code. 
And, while some reasoning functions of the proposed solutions may be similar to those carried out by expert human engineers and domain professionals, we focus on enabling presently impractical or impossible analyses, and much less on automation of manual tasks. 

\subsection{The Intelligent Development Aide}

The \emph{Intelligent Development Aide} (IDA) is a shared layer of services that offers the following to the overall solution: (i) intelligence, including learning, inference, and generative 
abilities; (ii) NL-based interaction; and, (iii)  general world knowledge and certain domain-specific expertise. 
The IDA will rely on present and future technologies that come under the umbrella of  AI, Generative AI, Machine Learning, Deep Learning, Large Language Models (LLMs), etc.   
It will be constructed, among other things, by  fine-tuning, enhancing and extending AI-based tools, relying on techniques like those of ~\cite{minaee2024LLMs,ding2023parameterEfficientFineTuning,shahaf2023ConceptAwareLLM,tamariShani2020NLisExecutableECL,rumpe2024NLtoModel} and future emerging ones.
With inputs from specifications of diverse systems, with textual and visual depictions of normal and faulty execution scenarios, the IDA will be trained to recognize unique software and system engineering issues and new delicate kinds of interdependencies.

\subsection{The i-model}

We introduce a new kind of model, termed \emph{i-model}, which offers fresh perspectives on some common modeling maxims:  
 
First, while precision is commonly needed to ensure correct system implementation, i-models will take advantage of what may appear quite the opposite. They will retain within model entities the expressive power of NL,  which includes sensitivity to context, flexible abstraction, generalization, associations, etc. Simulation and analysis tools will then rely on deferred -– \emph{just-in-time} (JIT) -– interpretation to endow NL and NL-like behavioral specifications with concrete meaning, aligned with the intended context, and abstraction level. 
 
Second, while logical flow and organization are essential to engineering, conceptual abstractions may not always lend themselves to being so depicted. For example, consider the difficulty in modeling a complex network of multiple class inheritances,  combined with natural language ambiguity, where, for instance, the word stop could mean a condition of no motion at all, or a process of slowing down to reach that condition, or the action of beginning to press the brake in order to begin this process, etc. 
In contrast, i-models will accommodate coexistence of multiple, diverse, non-hierarchical, overlapping and dynamic abstraction lattices.  
 
And third, modularity, encapsulation, and logical decomposition are central principles in software engineering, and in engineering in general. However, separately specified requirements often have tacit, unstated dependencies, which show up as exceptions, priorities, alternatives, complementary or concurrent actions, mutually exclusive conditions, etc. It
is commonly 
up to the engineers to infer these implicit relations, and to reflect their understanding in the implementation. In our automated construction of i-models from a wide range of specifications, special focus will be put on discovering such unstated relationships and capturing them in the model, despite the entanglement that they may imply. 

The i-models will store diverse information, including requirements, goals, behaviors, scenarios, emergent properties, etc., as well as groupings, abstractions, and relations of such entities. It will also contain meta information about potential changes due to evolution of the system and its environment, allowing further analysis of potential future trajectories. 

Finally,  the i-model will support \emph{unmodeling}; i.e., capturing entities and assumptions that should be ignored during execution and analysis, as well as operational environments in which the \srs is not expected to operate. Unmodeling will complement capabilities in existing modeling techniques to specify the exact intended operational design domain (ODD), directing IDA-based tools where to apply and where not to apply their vast and important knowledge~\cite{marronEtAl2023challengesInModelingMWCCIS}.

The immense knowledge stored in the i-model will be divided among three realms: (i) the entities themselves, including structured data and unstructured NL documents; (ii) the  relationships between entities, represented in the i-model database; and, (iii) the general and application-specific knowledge captured in the IDA components, both in advance, and following the building and analysis of a given i-model.

\subsection{The i-model builder} \label{sec:iModelBuilder}
 
Inputs to i-model building will include:  requirements documents,  specifications of reusable components~\cite{benvenisteHenzingerDamm2018contracts},
manual risk analyses~\cite{bjerga2016uncertaintyRiskAnalysis,haimes2018riskAnalysis},
whole models in various modeling languages, program code, 
documentation, example run logs of early prototypes,  
test cases, etc. 
Additional information, corrections and guidance provided interactively by engineers during model building will also be retained. Furthermore, the i-model builder can initiate queries, soliciting the engineers for missing information or confirmation of inferences. 

Beyond the now increasingly common translation of NL specifications into basic object models and computer programs, a unique feature of the i-model builder will be the automated, and optionally interactive, discovery and recording of undocumented tacit interdependencies among separately specified entities. For example, consider separately specified rules for an autonomous vehicle (AV) that may cause the AV to accelerate, as when entering a highway, or when instructed to follow another vehicle, or when returning to normal speed after a temporary slow down; consider a second set of rules specifying maximum legal speed and maximum recommended speed under certain conditions. 
The fact that the rules in the second set constrain or may  be in conflict with rules in the first set, will be captured at model-building time. When the effect is clear, i.e., that one rule takes priority over the other, this explicit specification will be generated automatically up front (where today it may be left as an  implementation detail); when the relation is in question, (say, what to do if the leader of a convoy exceeds the legal speed limit), 
the i-model builder will consult  the engineers.  

The builder will provide succinct summaries of the input information, and elaborate on its inferences applying logic and domain knowledge.  The added information  will also be stored in the i-model. For example, in our experiments with an LLM in creating a model from a description of a traffic scenario, the LLM added pedestrian objects, which were absent from the original requirements. 

The i-model will use a rich modern database to store the structured and unstructured information and the associated connections and relationships. As this structure will likely be too complex for humans to navigate and maintain directly, it will be supported by a multi-view presentation (see, e.g., \emph{slices} in~\cite{BriandCoq2012TraceabilityViewsSlices}), \emph{projecting} requested information as stand alone succinct text descriptions and dynamically created diagrams (e.g., Statecharts). Since many entities will already be in NL, the projection itself will be intuitive, but may still require other components for confirmed interpretation. Once confirmed, such projections can also serve as concise memory, to be fed back to the i-model.  

We expect the construction of the i-model builder to extend the growing amount of research on using GAI and LLMs for creating models from NL inputs;  see, e.g. ~\cite{ramackersChaudron2021NLtoUML,rumpe2024NLtoModel,WeissElYasafYaacov2024NLtoBP,KatzMarronSzekelyHarel2024MWBPCHATGPT,cosler2023nl2spec,aalst2024processModelingLLM}.

\subsection{The i-model execution and simulation engine}\label{sec:exec}

This component performs dynamic, deferred, just-in-time (JIT) interpretation of the definite and the NL information in the i-model, and outputs  descriptions of possible runs and the unfolding of
``what if'' scenarios at desired abstraction levels, as narratives, event logs, and even synthetic videos.

For illustration, consider the following example, which we ran  on OpenAI ChatGPT.
Prompt (slightly edited): \textit{``Consider a 3 lane street. 30 equal-size passenger cars are filling the street, standing in front of a traffic light. No cars arrive from the back of the queue throughout this scenario. The third car stalls and cannot move. No one stops to help. Describe the behaviors and positions of the 30 cars after the light turns green. Assume the light is green for a very long time.''} 

Here are a few excerpts from the LLM's response: \textit{``When the traffic light turns green:...\textbf{1.~Cars in the Middle Lane...:} (a) The first two cars in the middle lane (Cars~1 and~2) will start moving forward as the light turns green. (b) The third car in the middle lane (Car~3) will not move...(c) The cars behind the third car...will initially wait, expecting the third car to move. After realizing that it is not moving, they will either:
(i) Start honking,...; (ii) Some drivers may try to change lanes...(iii) Some drivers may remain stuck behind the third car if changing lanes is not possible...
\textbf{2.~Cars in the Left Lane...and Right Lane...:} 
(a) The cars in these lanes...will begin moving forward... (b) The first few cars...positioned parallel to the stationary third car, will drive past it without much issue. (c) The cars behind them will continue moving forward, potentially picking up speed...''}

When we divided the road into small virtual cells, the LLM was able to describe car movements through the numbered cells, making the scenario more mathematically analyzable. 

The LLM responses also contained errors; for example, they suggested that cars blocked behind the stalled car may not be able to change lanes at all, ignoring the specification that the traffic light stays green indefinitely, and no new cars arrive during the scenario;  
however, we believe that with proper training, responses can be made highly reliable. And in any case, the answers can be checked and validated interactively, and domain-specific knowledge can be added.
Note how the LLM incorporated into the responses its domain knowledge; for example,  adding  the possibilities of changing lanes and honking, which were not in the original specification. 

The execution engine will, of course, benefit from state-of-the-art execution and simulation techniques, like those in SysML, Statecharts, and  Scenario-based programming\cite{harelMarelly2003comeLetsPlay, HarelMarronWeiss2012BPCACM}, or in direct execution of NL specifications as in~\cite{tamariShani2020NLisExecutableECL} and references therein. Borrowing from techniques for test-case generation~\cite{wang2024softwareTestLLMSurvey}, the execution engine will also generate and process batches of diverse, yet relevant, ``what if'' scenarios, and store their execution results for further processing. 

\subsection{The i-model analysis engine}\label{sec:analysis}

The i-model analyzer will carry out the equivalent of formal model-checking, 
searching --- proactively --- for execution trajectories that lead to fault states. Treating the model as an NL-enriched graph, it will traverse its paths, interpreting entities and relationships subject to general and domain knowledge, including causalities, interdependencies and risks, while abiding by \emph{unmodeling} --- knowing what to exclude. In addition, the analyzer will offer query capabilities, e.g.,  for investigating complex scenarios, or the many connections of a given entity. It will also interface with classical model checkers and satisfiability modulo theory (SMT) constraint solvers for inspecting well-structured subsets and projections of the i-model, and for presenting the answers back in NL. 

For example, we described to ChatGPT two parallel synchronous state machines. With some trial and error, and with checking and corrections by engineers, the LLM was able to answer whether certain composite states were reachable or not, describe relevant paths, and construct the full state graph of the composite machine with all composite states and transitions. 
Other analytic LLM capabilities that can support i-model analysis are described in papers like~\cite{KatzMarronSzekelyHarel2024MWBPCHATGPT,sultan2024aiSysML}. 
These include computing when two independent periodic events may occur simultaneously, explaining behavior, articulating system properties, checking model consistency, 
etc. 
Furthermore, it is expected that LLM general and domain-specific analytical capabilities  will be extended and deepened, and they are likely to be intertwined with ongoing research in software and system engineering. 
Developments along the present roadmap can incorporate such enhancements, and target them specifically at early fault-detection. 

Still, automated validation techniques must be researched and developed. That is, assume that the fault-detection solution is confirmed to work well on reasonably tractable models, like compositions of small specifications. 
Can one trust the solution's answers on larger problems? 
And wouldn't a trusted automated external validation tool make the AI-based solution unnecessary altogether? We believe that with a combination of well-documented abstraction relations, AI-explainability, randomized testing of model answers, and powerful projection of relevant model perspectives, one can create high confidence in the model's answers.  

\subsection{Emergent Effect Detector} 

This component accepts outputs of system simulations, looking for expected and unexpected  patterns and emergent effects, both structural/spatial and behavioral/temporal.  Such effects may be previously specified as desired, undesired, or perhaps acceptable,  or they may require assessment. The tool will rely on the immense body of work in recognizing patterns, emergent effects, anomalies, etc., in sequential
data, like discrete event logs or continuous signals, and in spatial and structural information, like images and videos. See, e.g., ~\cite{pang2021deepLearningAnomalyRec,fieguth2022MLandPatternRecognition,noering2021patternDIscoveryAutoENcoder,nickovic2022surveySTLspecMining}. 
The results will be presented formally and in NL for manual and automated analysis. 

\subsection{Repair Advisor}
 
The i-model's sheer size may interfere with its maintenance, calling for a repair advisor that accepts a description of an issue and proposes changes to the system or to its technical and physical environment. A key distinction from common program repair~\cite{Zhang2023AutomaticProgramRepairByMLSurvey} is the primary focus on pinpointing the model components that should be changed and on describing the ensuing impact on system behavior, while the technical details of the actual change are secondary.

\section{Conclusion}
We are now  in the process of initiating a research project following the above roadmap. Development of models and tools that enable simulation and analysis of highly complex systems based only on early specifications can dramatically enhance our ability to develop reliable, safe, and productive super-reactive systems. A combination of the recent advances in AI and a fresh perspective on what may and may not qualify as a model entity or be acceptable as a simulation result, may enable the achievement of this tantalizing goal.

\section*{Acknowledgments}
This research was funded in part by an NSFC-ISF grant to DH,  issued jointly by the  National Natural Science Foundation of China (NSFC) and the Israel Science Foundation (ISF grant 3698/21). Additional support was provided by a research grant to DH from Louis J. Lavigne and Nancy Rothman, the Carter Chapman Shreve Family Foundation, Dr. and Mrs. Donald Rivin, and the Estate of Smigel Trust.
\bibliographystyle{ieeetr}

\begin{thebibliography}{10}

\bibitem{harelPnueli1984ReactiveSystems}
D.~Harel and A.~Pnueli, ``On the development of reactive systems,'' in {\em Logics and models of concur. sys.}, pp.~477--498, Springer, 1984.

\bibitem{franceRumpe2007MDEresearchroadmap}
R.~France and B.~Rumpe, ``Model-driven development of complex software: A research roadmap,'' in {\em Future of Software Engineering (FOSE'07)}, pp.~37--54, IEEE, 2007.

\bibitem{laplante2022SWEngineering}
P.~A. Laplante and M.~Kassab, {\em What every engineer should know about software engineering}.
\newblock CRC Press, 2022.

\bibitem{palanque2017SOTAinFM}
R.~Oliveira, P.~Palanque, B.~Weyers, J.~Bowen, and A.~Dix, ``State of the art on formal methods for interactive systems,'' {\em The handbook of formal methods in human-computer interaction}, pp.~3--55, 2017.

\bibitem{fremontVincentelli2023scenic}
D.~J. Fremont, E.~Kim, T.~Dreossi, S.~Ghosh, X.~Yue, A.~L. Sangiovanni-Vincentelli, and S.~A. Seshia, ``Scenic: A language for scenario specification and data generation,'' {\em Machine Learning}, vol.~112, no.~10, 2023.

\bibitem{saqui2021ModelCheckingSysMLCockpitDoor}
P.~de~Saqui-Sannes, L.~Apvrille, and R.~Vingerhoeds, ``Checking {SysML} models against safety and security properties,'' {\em Journal of Aerospace Information Systems}, vol.~18, no.~12, pp.~906--918, 2021.

\bibitem{zahid2022systematicMappingFMinRE}
F.~Zahid, A.~Tanveer, M.~M. Kuo, and R.~Sinha, ``A systematic mapping of semi-formal and formal methods in requirements engineering of industrial cyber-physical systems,'' {\em J. of Intel. Mfg.}, vol.~33, no.~6, 2022.

\bibitem{weyersPalanque2017handbook}
B.~Weyers, J.~Bowen, A.~Dix, and P.~Palanque, {\em The handbook of formal methods in human-computer interaction}.
\newblock Springer, 2017.

\bibitem{huangMcGinnis2020verifyingSysMLPetriNets}
E.~Huang, L.~F. McGinnis, and S.~W. Mitchell, ``Verifying {SysML} activity diagrams using formal transformation to petri nets,'' {\em Systems Engineering}, vol.~23, no.~1, pp.~118--135, 2020.

\bibitem{Rahim2021HierachicalpETRINETSsysmlVerActivityReqRelations}
M.~Rahim, M.~Boukala-Ioualalen, and A.~Hammad, ``Hierarchical colored {Petri nets} for the verification of {SysML} designs-activity-based slicing approach,'' in {\em Advances in Computing Systems and Applications: Proc. 4th Conf. on Comp. Sys. and App.}, pp.~131--142, Springer, 2021.

\bibitem{Ghezzi2020earlyValidationMultiConcerns}
N.~Li, C.~Tsigkanos, Z.~Jin, Z.~Hu, and C.~Ghezzi, ``Early validation of cyber--physical space systems via multi-concerns integration,'' {\em Journal of Systems and Software}, vol.~170, p.~110742, 2020.

\bibitem{harelMarronMizrahiWeissKatzKantor2013compositional}
D.~Harel, A.~Kantor, G.~Katz, A.~Marron, L.~Mizrahi, and G.~Weiss, ``On composing and proving the correctness of reactive behavior,'' in {\em EMSOFT 2013}, pp.~1--10, IEEE, 2013.

\bibitem{Lattimore2022EuropaSalnderSImulation}
M.~Lattimore, R.~Karban, M.~P. Gomez, E.~Bovre, and G.~E. Reeves, ``A model-based approach for {Europa} lander mission concept exploration,'' in {\em 2022 IEEE Aerospace Conference (AERO)}, pp.~1--13, IEEE, 2022.

\bibitem{gorecki2019riskModelingIndustrial}
S.~Gorecki, J.~Ribault, G.~Zacharewicz, Y.~Ducq, and N.~Perry, ``Risk management and distributed simulation in papyrus tool for decision making in industrial context,'' {\em Comput. \& Indus. Engineering}, vol.~137, 2019.

\bibitem{lo2021digitalTwinInDev}
C.~Lo, C.-H. Chen, and R.~Y. Zhong, ``A review of digital twin in product design and development,'' {\em Adv. Eng. Informatics}, vol.~48, p.~101297, 2021.

\bibitem{dagstuhl2024AVsafety}
J.~Deshmukh, B.~K\"{o}nighofer, D.~Ni\v{c}kovi\'{c}, and F.~Cano, ``{Safety Assurance for Autonomous Mobility (Dagstuhl Seminar 24071)},'' {\em Dagstuhl Reports}, vol.~14, no.~2, pp.~95--119, 2024.

\bibitem{cederbladh2024earlyVandVofModels}
J.~Cederbladh, A.~Cicchetti, and J.~Suryadevara, ``Early validation and verification of system behaviour in model-based systems engineering: a systematic literature review,'' {\em ACM Transactions on Software Engineering and Methodology}, vol.~33, no.~3, pp.~1--67, 2024.

\bibitem{horvath2023pragmaticVerificationWIthExecutableSysML}
B.~Horv{\'a}th, V.~Moln{\'a}r, B.~Graics, {\'A}.~Hajdu, I.~R{\'a}th, {\'A}.~Horv{\'a}th, R.~Karban, G.~Trancho, and Z.~Micskei, ``Pragmatic verification and validation of industrial executable {SysML} models,'' {\em Systems Engineering}, vol.~26, no.~6, pp.~693--714, 2023.

\bibitem{harelSifakisMarron2020autonomicsPNAS}
D.~Harel, A.~Marron, and J.~Sifakis, ``Autonomics: In search of a foundation for next-generation autonomous systems,'' {\em Proceedings of the National Academy of Sciences}, vol.~117, no.~30, pp.~17491--17498, 2020.

\bibitem{EdwardLee2024certaintyOrIntelligence}
E.~A. Lee, ``Certainty or intelligence: Pick one!,'' in {\em Design, Automation \& Test in Europe (DATE)}, pp.~1--2, IEEE, 2024.

\bibitem{AIESE2024Conf}
{AIESE}, ``{15th Int. Conf. on AI-empowered Software Engineering – AIESE 2024 (Formerly JCKBSE)},'' 2024.
\newblock \\ \url{https://easyconferences.eu/aiese2024/}; Accessed Aug. 2024.

\bibitem{clarke2003counterexamples}
E.~Clarke and H.~Veith, {\em Counterexamples revisited: Principles, algorithms, applications}.
\newblock Springer, 2003.

\bibitem{seippHelmert2018CegarAndAbstractionTypesPlanning}
J.~Seipp and M.~Helmert, ``Counterexample-guided cartesian abstraction refinement for classical planning,'' {\em J. of Artificial Intel. Res.}, vol.~62, 2018.

\bibitem{minaee2024LLMs}
S.~Minaee, T.~Mikolov, N.~Nikzad, M.~Chenaghlu, R.~Socher, X.~Amatriain, and J.~Gao, ``Large language models: A survey,'' {\em arXiv preprint arXiv:2402.06196}, 2024.

\bibitem{ding2023parameterEfficientFineTuning}
N.~Ding, Y.~Qin, G.~Yang, F.~Wei, Z.~Yang, Y.~Su, S.~Hu, Y.~Chen, C.-M. Chan, W.~Chen, {\em et~al.}, ``Parameter-efficient fine-tuning of large-scale pre-trained language models,'' {\em Nature Mach. Intel.}, vol.~5, no.~3, pp.~220--235, 2023.

\bibitem{shahaf2023ConceptAwareLLM}
C.~Shani, J.~Vreeken, and D.~Shahaf, ``Towards concept-aware large language models,'' {\em arXiv preprint arXiv:2311.01866}, 2023.

\bibitem{tamariShani2020NLisExecutableECL}
R.~Tamari, C.~Shani, T.~Hope, M.~R.~L. Petruck, O.~Abend, and D.~Shahaf, ``{L}anguage (re)modelling: {T}owards embodied language understanding,'' in {\em Proc. of the 58th Annual Meeting of the ACL} (D.~Jurafsky, J.~Chai, N.~Schluter, and J.~Tetreault, eds.), ACL, 2020.

\bibitem{rumpe2024NLtoModel}
L.~Netz, J.~Michael, and B.~Rumpe, ``From natural language to web applications: Using large language models for model-driven software engineering,'' in {\em Modellierung 2024}, pp.~179--195, Gesellschaft f{\"u}r Informatik eV, 2024.

\bibitem{marronEtAl2023challengesInModelingMWCCIS}
A.~Marron, I.~R. Cohen, G.~Frankel, D.~Harel, and S.~Szekely, ``Challenges in modeling and unmodeling complex reactive systems: Interaction networks, reaction to emergent effects, reactive rule composition, and multiple time scales,'' {\em Springer CCIS}, 2024.

\bibitem{benvenisteHenzingerDamm2018contracts}
A.~Benveniste, B.~Caillaud, D.~Nickovic, R.~Passerone, J.-B. Raclet, P.~Reinkemeier, A.~Sangiovanni-Vincentelli, W.~Damm, T.~A. Henzinger, K.~G. Larsen, {\em et~al.}, ``Contracts for system design,'' {\em Foundations and Trends{\textregistered} in Electronic Design Automation}, vol.~12, no.~2-3, pp.~124--400, 2018.

\bibitem{bjerga2016uncertaintyRiskAnalysis}
T.~Bjerga, T.~Aven, and E.~Zio, ``Uncertainty treatment in risk analysis of complex systems: The cases of stamp and fram,'' {\em Reliability Engineering \& System Safety}, vol.~156, pp.~203--209, 2016.

\bibitem{haimes2018riskAnalysis}
Y.~Y. Haimes, ``Risk modeling of interdependent complex systems of systems: Theory and practice,'' {\em Risk Analysis}, vol.~38, no.~1, pp.~84--98, 2018.

\bibitem{BriandCoq2012TraceabilityViewsSlices}
S.~Nejati, M.~Sabetzadeh, D.~Falessi, L.~Briand, and T.~Coq, ``A {SysML}-based approach to traceability management and design slicing in support of safety certification: Framework, tool support, and case studies,'' {\em Information and Software Technology}, vol.~54, no.~6, pp.~569--590, 2012.

\bibitem{ramackersChaudron2021NLtoUML}
G.~J. Ramackers, P.~P. Griffioen, M.~B. Schouten, and M.~R. Chaudron, ``From prose to prototype: synthesising executable {UML} models from natural language,'' in {\em MODELS-C}, pp.~380--389, IEEE, 2021.

\bibitem{WeissElYasafYaacov2024NLtoBP}
T.~Yaacov, A.~Elyasaf, and G.~Weiss, ``{Boosting LLM-Based Software Generation by Aligning Code with Requirements},'' in {\em Proc. 14th Int. Model-Driven Requirements Engineering Workshop (MoDRE)}, 2024.

\bibitem{KatzMarronSzekelyHarel2024MWBPCHATGPT}
D.~Harel., G.~Katz., A.~Marron., and S.~Szekely., ``On augmenting scenario-based modeling with generative {AI},'' in {\em MODELSWARD 2024}, pp.~235--246, 2024.

\bibitem{cosler2023nl2spec}
M.~Cosler, C.~Hahn, D.~Mendoza, F.~Schmitt, and C.~Trippel, ``nl2spec: interactively translating unstructured natural language to temporal logics with large language models,'' in {\em CAV}, pp.~383--396, Springer, 2023.

\bibitem{aalst2024processModelingLLM}
H.~Kourani, A.~Berti, D.~Schuster, and W.~M. van~der Aalst, ``Process modeling with large language models,'' in {\em International Conference on Business Process Modeling, Development and Support}, pp.~229--244, Springer, 2024.

\bibitem{harelMarelly2003comeLetsPlay}
D.~Harel and R.~Marelly, {\em Come, let's play: scenario-based programming using LSCs and the play-engine}, vol.~1.
\newblock Springer, 2003.

\bibitem{HarelMarronWeiss2012BPCACM}
D.~Harel, A.~Marron, and G.~Weiss, ``Behavioral programming,'' {\em Communications of the ACM}, vol.~55, no.~7, pp.~90--100, 2012.

\bibitem{wang2024softwareTestLLMSurvey}
J.~Wang, Y.~Huang, C.~Chen, Z.~Liu, S.~Wang, and Q.~Wang, ``Software testing with large language models: Survey, landscape, and vision,'' {\em IEEE Transactions on Software Engineering}, 2024.

\bibitem{sultan2024aiSysML}
B.~Sultan and L.~Apvrille, ``Ai-driven consistency of sysml diagrams,'' in {\em Proceedings of the ACM/IEEE 27th International Conference on Model Driven Engineering Languages and Systems}, pp.~149--159, 2024.

\bibitem{pang2021deepLearningAnomalyRec}
G.~Pang, C.~Shen, L.~Cao, and A.~V.~D. Hengel, ``Deep learning for anomaly detection: A review,'' {\em ACM comput. surv.}, vol.~54, no.~2, pp.~1--38, 2021.

\bibitem{fieguth2022MLandPatternRecognition}
P.~Fieguth, {\em An introduction to pattern recognition and machine learning}.
\newblock Springer, 2022.

\bibitem{noering2021patternDIscoveryAutoENcoder}
F.~K.-D. Noering, Y.~Schroeder, K.~Jonas, and F.~Klawonn, ``Pattern discovery in time series using autoencoder in comparison to nonlearning approaches,'' {\em Integrated Computer-Aided Engineering}, vol.~28, no.~3, 2021.

\bibitem{nickovic2022surveySTLspecMining}
E.~Bartocci, C.~Mateis, E.~Nesterini, and D.~Nickovic, ``Survey on mining signal temporal logic specifications,'' {\em Information and Computation}, vol.~289, p.~104957, 2022.

\bibitem{Zhang2023AutomaticProgramRepairByMLSurvey}
Q.~Zhang, C.~Fang, Y.~Ma, W.~Sun, and Z.~Chen, ``A survey of learning-based automated program repair,'' {\em ACM Transactions on Software Engineering and Methodology}, vol.~33, no.~2, pp.~1--69, 2023.

\end{thebibliography}

\end{document}